\documentclass[notoc]{JHEP3}
\usepackage{amsmath}
\usepackage{amssymb}
\usepackage{amsfonts}
\usepackage{amscd}
\usepackage{bbm} 
\usepackage[footnotesize]{caption}
\usepackage{graphicx} 
\usepackage{braket} 
\usepackage{mathrsfs}
\usepackage{BetterCite}
\usepackage[center,footnotesize,hang]{subfigure}
\usepackage{HEP5} 
\usepackage[bbgreekl]{MyBbol}
\usepackage{longtable}

\def\D{\mathrm{d}} 

\def\Tr{\text{Tr}}

\def\ChargeC{\mathrm{C}}		
 
\newcommand{\SFConjugate}[1]{{#1}^\ChargeC} 
\newcommand{\SuperField}[1]{\bbsymbol{#1}}

\abstract{
We present a construction kit for calculating two-loop
beta functions in $N\!=\!1$ supersymmetric theories
for the operators of the superpotential using supergraph techniques.
In particular, it allows to compute the beta functions for every desired,
even higher dimensional, operator of the superpotential from the 
wavefunction renormalization constants of the theory.
We apply this method to calculate the two-loop beta functions for the
lowest-dimensional effective neutrino mass operator in the 
Minimal Supersymmetric Standard Model (MSSM) 
and for the Yukawa couplings in the MSSM extended 
by singlet superfields and the mass matrix for the latter.
Our method can be applied to any $N\!=\!1$ supersymmetric theory.
}
\title{
Supergraph Techniques and 
Two-Loop Beta-Functions for
Renormalizable and Non-Renormalizable
Operators
}

\author{Stefan Antusch\\
Physik-Department T30, 
Technische Universit\"{a}t M\"{u}nchen\\ 
James-Franck-Stra{\ss}e,
85748 Garching, Germany\\
E-mail: \email{santusch@ph.tum.de} 
}
\author{Michael Ratz\\
Physik-Department T30, 
Technische Universit\"{a}t M\"{u}nchen\\ 
James-Franck-Stra{\ss}e,
85748 Garching, Germany\\
E-mail: \email{mratz@ph.tum.de} 
}

\keywords{Renormalization Group Equation, Supersymmetry, Neutrino Mass}
\preprint{TUM-HEP-453/02}
\begin{document}





\maketitle

\newpage
\section{Introduction}

In order to compare experimental results with predictions from models 
beyond the Standard Model (SM), like unified theories, 
it is essential to evolve the parameters of the models 
from high to low energies. 
This is accomplished with the renormalization group equations (RGE's)
for the operators in the theory. 
Besides the renormalizable operators there can be higher dimensional, 
non-renormalizable operators if the theory is considered as effective. 

In a previous study \cite{Antusch:2001ck}, 
we derived a general method for calculating 
\(\beta\)-functions from counterterms 
in MS-like renormalization schemes, which works for tensorial quantities. 
It is simplified considerably in supersymmetric (SUSY) theories,
since due to the non-renormalization theorem \cite{Wess:1974kz,Iliopoulos:1974zv} 
only wavefunction renormalization has to be considered
for operators of the superpotential.  
However, in a component field description, 
no use can be made of the 
theorem with respect to 
gauge loop corrections since it is no longer manifest when a 
supergauge, as for example Wess-Zumino-gauge, has been fixed.
The supergraph technique \cite{Delbourgo:1975jg,Salam:1975pp,Fujikawa:1975ay,Grisaru:1979wc},
on the other hand, allows to use the 
non-renormalization theorem, since SUSY is kept manifest. 
Moreover, it has the advantage that the number of independent
diagrams is clearly reduced compared to the component field calculations.

We therefore present a method to calculate \(\beta\)-functions in 
supersymmetric theories for operators of the superpotential from wavefunction
renormalization. These operators may be non-renormalizable
since for the latter the non-renormalization theorem holds as well
\cite{Weinberg:1998uv}, and they do not affect the wavefunction renormalization
constants in leading order in an effective field theory expansion.
As an application, we consider the Minimal Supersymmetric Standard Model
(MSSM) extended by 
singlet superfields, which contain 
right-handed neutrinos relevant for models of neutrino mass. 
We compute and specify the wavefunction 
renormalization constants. From these, the two loop RGE's for
the Yukawa couplings and for a possible 
mass matrix for the singlet neutrino superfields 
are obtained by the supergraph method. 
The technique can be used to compute the two loop RGE's for every desired 
higher dimensional operator of the superpotential, since the 
non-renormalization theorem guarantees that no vertex corrections 
contribute.
Furthermore, we consider the lowest dimensional neutrino mass operator. 
Its RGE 
is known at the one loop level for the SM 
\cite{Chankowski:1993tx,Babu:1993qv,Antusch:2001ck}, Two Higgs Doublet Models 
\cite{Babu:1993qv,Antusch:2002vn} and for the MSSM 
\cite{Chankowski:1993tx,Babu:1993qv,Antusch:2002vn}.
With the supergraph method we calculate the two 
loop RGE for the neutrino mass operator 
from the MSSM wavefunction renormalization constants.

\clearpage
\section{Review of the Supergraph Method}
 
Consider a general supersymmetric gauge theory with a gauge part 
described by the usual Lagrangian
\begin{eqnarray}
 \mathscr{L}
 & =& 
 \int\D^4\theta\,
 \sum_{i,j=1}^{N_\SuperField{\Phi}}
 \overline{\SuperField{\Phi}}^{(i)}\,
 	\left[\exp(2g\cdot\SuperField{V})\right]_{ij}\,
 \SuperField{\Phi}^{(j)}
 +\left[\frac{1}{4}\int\D^2\theta\,\sum_{n=1}^S
 	\SuperField{W}_\alpha^n\SuperField{W}^{n\,\alpha}
		+\text{h.c.}\right]
 \nonumber\\
 &&
 +\left[\int\D^2 \theta\,\mathscr{W}	+\text{h.c.}\right]
 +\mathscr{L}_\mathrm{Ghost}+\mathscr{L}_\mathrm{Gauge\:Fixing}
\end{eqnarray}
where
\begin{subequations} 
\begin{eqnarray}
 \SuperField{W}_\alpha^n & = & \frac{1}{8g_n}\,\overline{\boldsymbol{D}}^2
 	\left[\exp(2g_n\,\SuperField{V}^n)\,\boldsymbol{D}_\alpha\,
	\exp(-2g_n\SuperField{V}^n)\right]
 \;,\\
 g\cdot\SuperField{V} 
 & := & 
 \sum_{n=1}^S g_n\,\SuperField{V}^n
 \quad \text{and}\quad
 \SuperField{V}^n=\sum_{A=1}^{\dim G_n} \SuperField{V}_n^A\,\mathsf{T}_n^A\;.
\end{eqnarray}
\end{subequations}
The renormalizable part of the superpotential reads
\begin{eqnarray}\label{eq:SuperpotentialWithoutHOPs}
 \mathscr{W}_\mathrm{ren} = \frac{1}{6}\sum_{i,j,k=1}^{N_\SuperField{\Phi}}
 \lambda_{(ijk)}\,\SuperField{\Phi}^{(i)}\,
 	\SuperField{\Phi}^{(j)}\,\SuperField{\Phi}^{(k)}
 \;.
\end{eqnarray}
Possible mass terms are ignored for the present as they do not affect the 
$\beta$-functions of the model.
The \(N_\SuperField{\Phi}\) superfields \(\SuperField{\Phi}^{(i)}\) transform 
under the irreducible representations (irreps)
\(R^{(i)}_1\times\cdots\times R^{(i)}_S\) of the gauge group
\(G_1\otimes\cdots\otimes G_S\). 
\(\{(\mathsf{T}_n^A)_{i}^{j}\}_{A=1}^{\dim G_n}\)
denote the generators of $G_n$. 
The indices $i,j,\dots$ run over all irreps, families and the
representation space.

\subsection{Wavefunction Renormalization Constants}

Due to the non-renormalization theorem, the RGE's for operators of the 
superpotential are governed by the wavefunction renormalization constants for
the  superfields 
${Z}_{ij} =  \mathbbm{1}_{ij} + \delta Z_{ij} $ 
which relate the bare $\SuperField{\Phi}^{(i)}_{\mathrm{B}}$ 
and the renormalized superfields,
\begin{eqnarray}
\SuperField{\Phi}^{(i)}_{\mathrm{B}} = \sum_{j=1}^{N_\SuperField{\Phi}} 
Z_{ij}^{\frac{1}{2}} \SuperField{\Phi}^{(j)}\; .
\end{eqnarray}
Results are obtained with dimensional regularization via dimensional 
reduction \cite{Siegel:1979wq,Capper:ns}. 
 At the one loop level and in \(d=4-\epsilon\) dimensions, 
\(\delta Z_{ij}\) is  given by
\begin{eqnarray}\label{eq:OneLoopDeltaZ}
 -\delta Z_{ij}^{(1)} = \frac{1}{(4\pi)^2}\frac{1}{\epsilon}\,
 \left[
 \sum_{k,\ell=1}^{N_\SuperField{\Phi}}
 \lambda_{ik\ell}^*\lambda_{jk\ell}
 	-4\sum_{n=1}^S g_n^2\,c_2\left(R^{(i)}_n\right)\,\delta_{ij}\right]\;.
\end{eqnarray}
In equation (\ref{eq:OneLoopDeltaZ}) and in the following, we use the group-theoretical 
constants
 \begin{subequations}\label{eq:GroupTheoreticalConstants}
 \begin{eqnarray}
  c_1(G)\,\delta^{AB} & := & \sum_{C,D}f^{ACD}f^B{}_{CD}\;,\\
  c_2(R)\,\delta_{ab} & := & \sum_A (\mathsf{T}^A\mathsf{T}^A)_{ab}\;,
  \label{eq:QuadraticCasimir}\\
  \ell(R)\,\delta^{AB} & := & \Tr(\mathsf{T}^A\mathsf{T}^B)\;,
  \label{eq:DynkinIndex}
 \end{eqnarray}
 \end{subequations}
 with the matrix representations \(\{\mathsf{T}^A\}_{A=1}^{\dim G}\) of 
 the generators of $G$ corresponding to the irrep \(R\)
 and the structure constants \(f^A{}_{BC}\). 
 \(\ell(R)\) is known as Dynkin index of the irrep \(R\) and
 \(c_2(R)\) as the quadratic Casimir.
 They are related by
 \begin{eqnarray}
  c_2(R)=\frac{\dim G}{\dim R}\,\ell(R)\;,
 \end{eqnarray}
 with \(\dim G\) and \(\dim R\) being the dimension of the group \(G\) 
 and the irrep \(R\), respectively.
 Often the generators of the irrep \(\boldsymbol{N}\) of \(\mathrm{SU}(N)\) are  
 normalized such that \(\ell(\boldsymbol{N})=\frac{1}{2}\) holds. 
 \(c_2\) can then be obtained via \(c_2(\boldsymbol{N})=\frac{N^2-1}{2N}\) 
 while for a \(\mathrm{U}(1)\) theory both \(\ell(R)\) and \(c_2(R)\) are 
 replaced by \(q^2\) where \(q\) is the \(\mathrm{U}(1)\) charge of 
 \(\SuperField{\Phi}\). 
 For any non-trivial irrep of  \(\text{SU}(N)\) the invariant 
 \(c_1(\boldsymbol{N})\) is given by \(N\).

On the two-loop level the renormalization group equations are determined from 
the formula \cite{West:1984dg}
\begin{eqnarray}
 -\delta Z_{ij}^{(2)}
 & = &
 \frac{-2+\epsilon}{(4\pi)^4\,\epsilon^2}
 \,\left[
  4\sum_{n,m=1}^Sg_n^2\,c_2\!\left(R^{(i)}_n\right)\,g_m^2
  	\,c_2\!\left(R^{(j)}_m\right)\,\delta_{ij}
 \right.\nonumber\\
 &&
 \hphantom{\frac{-2+\epsilon}{(4\pi)^4\,\epsilon^2}\,\left[\right.}
  +2\,\sum_{n=1}^S g_n^4\,c_2\!\left(R^{(i)}_n\right)\,\left(\overline{\ell}_{n}-3\,c_1(G_n)
  \right)\,\delta_{ij}
 \nonumber\\
 & & 
 \hphantom{\frac{-2+\epsilon}{(4\pi)^4\,\epsilon^2}\,\left[\right.}
  +\sum_{n=1}^S \sum_{k,\ell=1}^{N_\SuperField{\Phi}} g_n^2
  \left(-\,c_2\!\left(R^{(i)}_n\right)
  +2\,c_2\!\left(R^{(\ell)}_n\right)\right)\,\lambda_{ik\ell}^*\lambda_{jk\ell}
 \nonumber\\
 &&
 \left. 
 \hphantom{\frac{-2+\epsilon}{(4\pi)^4\,\epsilon^2}\,\left[\right.}
  -\frac{1}{2}\sum_{k,\ell,r,s,t=1}^{N_\SuperField{\Phi}}
  \lambda_{ik\ell}^*\lambda_{\ell st}\,\lambda_{rst}^*\lambda_{jkr}
 \right]\;.
 \label{eq:TwoLoopDeltaZ}
\end{eqnarray}
Here $\overline{\ell}_{n}$ is defined by
\begin{eqnarray}
 \overline{\ell}_{n}:=  \sum_{i=1}^{N_\SuperField{\Phi}}
 \frac{\ell\left(R_n^{(i)}\right)}{\dim\left(R_n^{(i)}\right)} \;.
\end{eqnarray}

In addition to the superpotential of equation 
(\ref{eq:SuperpotentialWithoutHOPs}), higher dimensional operators 
may appear in the superpotential of an effective theory.
These operators 
are generally suppressed by inverse powers of a large mass scale $M_X$.
Though these operators are non-renormalizable by power counting, 
in the effective field theory approach one can 
renormalize the theory in 
an expansion in inverse 
powers of $M_X$.
In the leading order of this expansion, 
the higher dimensional operators 
do not contribute to the wavefunction renormalization.

\begin{figure}[h]
 \begin{center}
  \subfigure[\label{subfig:OneLoop1}]{%
  \(\vcenter{\hbox{\includegraphics{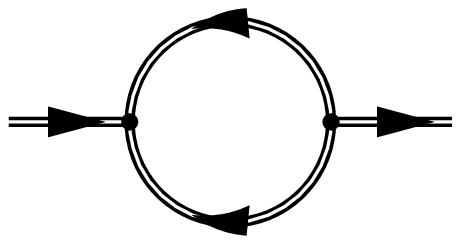}}}\)}
  \hfil
  \subfigure[\label{subfig:OneLoop2}]{%
  \(\vcenter{\hbox{\includegraphics{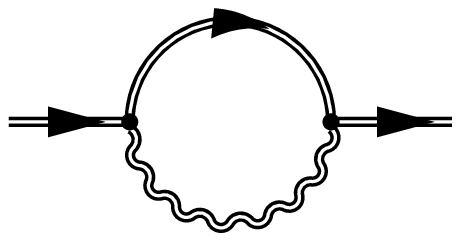}}}\)}
 \end{center}
 \caption{One-loop supergraphs which contribute to the 
 \(\overline{\SuperField{\Phi}}\SuperField{\Phi}\) propagator.
 \label{fig:OneLoopSelfEnergyDiagrams}}
\end{figure}

\begin{figure}[h]
 \begin{center}
  \subfigure[\label{subfig:TwoLoop1}]{%
  \(\vcenter{\hbox{\includegraphics{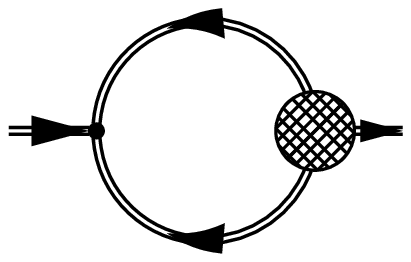}}}\)}
  \hfil
  \subfigure[\label{subfig:TwoLoop2}]{%
  \(\vcenter{\hbox{\includegraphics{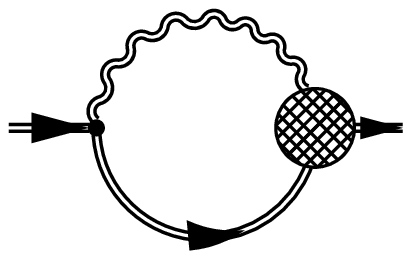}}}\)}
  \hfil
  \subfigure[\label{subfig:TwoLoop3}]{%
  \(\vcenter{\hbox{\includegraphics{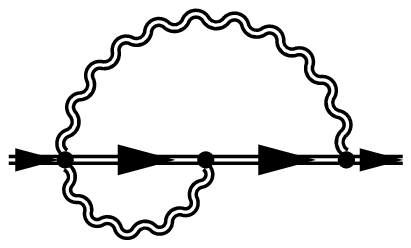}}}\)}
\\
  \subfigure[\label{subfig:TwoLoop4}]{%
  \(\vcenter{\hbox{\includegraphics{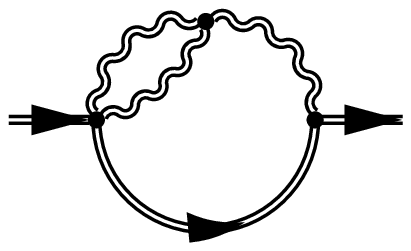}}}\)}
  \hfil
  \subfigure[\label{subfig:TwoLoop5}]{%
  \(\vcenter{\hbox{\includegraphics{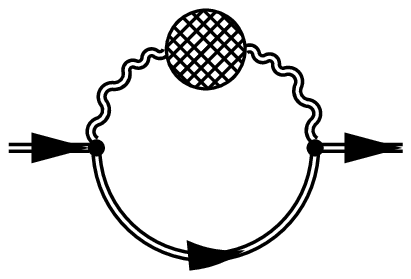}}}\)}
  \hfil
  \subfigure[\label{subfig:TwoLoop6}]{%
  \(\vcenter{\hbox{\includegraphics{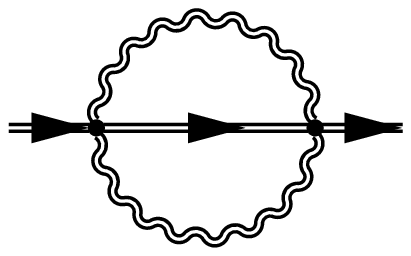}}}\)}
\\
  \subfigure[\label{subfig:TwoLoop7}]{%
  \(\vcenter{\hbox{\includegraphics{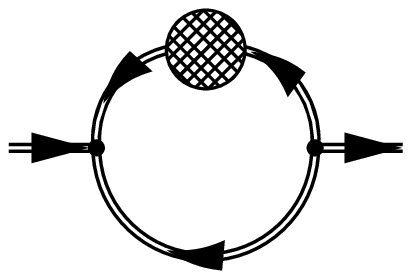}}}\)}
  \hfil
  \subfigure[\label{subfig:TwoLoop8}]{%
  \(\vcenter{\hbox{\includegraphics{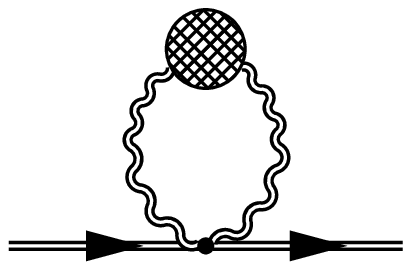}}}\)}
  \hfil
  \subfigure[\label{subfig:TwoLoop9}]{%
  \(\vcenter{\hbox{\includegraphics{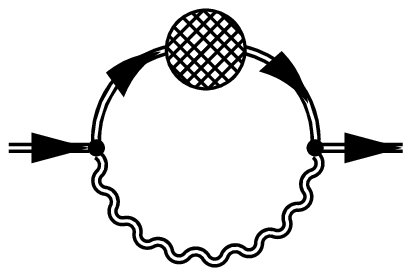}}}\)}
 \end{center}
 \caption{Two-loop supergraphs which contribute to the 
 \(\overline{\SuperField{\Phi}}\SuperField{\Phi}\) propagator. A blob denotes
 the relevant one-particle irreducible graph including any one-loop counterterm
 that may be required \cite{West:1984dg}.\label{fig:TwoLoopSelfEnergyDiagrams}}
\end{figure}

In supergraphs, we represent chiral superfields as straight double
lines while vector-superfields are indicated by wiggly double lines,
 \[
 \begin{array}{rcl}
 	\SuperField{\Phi} & : & 
	\vcenter{\hbox{\includegraphics{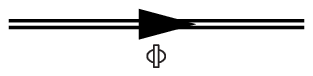}}}\;,
	\\
	\SuperField{V} & : & 
	\vcenter{\hbox{\includegraphics{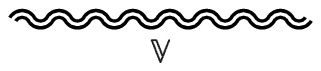}}}\;.
 \end{array}
 \]
The diagrams
relevant for the calculation of the
wavefunction renormalization constants for
the matter superfield,
equation \eqref{eq:OneLoopDeltaZ} and \eqref{eq:TwoLoopDeltaZ}, 
are shown in figure \ref{fig:OneLoopSelfEnergyDiagrams} and
\ref{fig:TwoLoopSelfEnergyDiagrams}, respectively.

\subsection{$\beta$-Functions from Wavefunction Renormalization Constants}

To calculate $\beta$-functions from the wavefunction renormalization 
constants, it is convenient to subdivide the general indices $\{i,j,\dots\}$ into
indices $\{r,s,\dots\}$ for the irreducible representations, $\{f,g,\dots\}$ for the particle 
families and $\{a,b,\dots\}$ for the representation space, i.e. $i=(r,f,a)$.
The wavefunction renormalization constants $Z_{ij}$ are diagonal
with respect to the representation and the representation space indices and are
matrices in flavour-space.  
In the rest of the paper we will write $Z_{ij}=Z_{r}$ and suppress 
flavour and representation space indices.

Due to the non-renormalization theorem, in a supersymmetric theory
a bare quantity \(Q_\mathrm{B}\) of the superpotential
and the corresponding renormalized one, \(Q\), are related by
\begin{eqnarray}\label{eq:QBare}
 Q_\mathrm{B} = 
 	\left(\prod\limits_{r\in I} Z_{r}^{n_r}\right)
	\,Q\,\mu^{D_Q\epsilon}\,
	\left(\prod\limits_{s\in J} Z_{s}^{n_s}\right)\; ,
\end{eqnarray}
where matrix multiplication with respect to flavour 
indices is implicit and 
the sets of superfield indices \(I=\{1,\dots M\}\) and  \(J=\{M+1,\dots N\}\) 
denote the
wavefunction renormalization constants multiplied from the left and the right
respectively.
$Q$ may correspond to a
renormalizable or non-renormalizable operator.
	
The wavefunction renormalization constants 
$Z_r$ can be expanded in $\epsilon$,
\begin{eqnarray}
 Z_{r}
 =
 1+\sum_{k\ge1} \frac{Z_{r,k}}{\epsilon^k}\;.
\end{eqnarray}
Following the steps of the derivation in \cite{Antusch:2001ck}, we find
\begin{eqnarray}\label{eq:BetaFunctionSusy}
 \beta_Q
 & = &
 Q\cdot\sum_{s\in J} n_s
        \left[\sum_A D_{V_A}\,\Braket{\frac{\D Z_{s,1}}{\D V_A}|V_A}\right]	
 \nonumber\\
 && \hphantom{Q\cdot\sum_{s\in J} n_s\left[\right.}
 +\sum_{r\in I}n_r
        \left[\sum_A D_{V_A}\,\Braket{\frac{\D Z_{r,1}}{\D V_A}|V_A}\right]
 			\cdot Q\;,
 \label{eq:SUSYBetaFunction}
\end{eqnarray}
where \(\{V_A\}\) 
denotes the set of all variables of the theory including
the variable \(Q\) under consideration and \(\beta_Q\) is the usual 
\(\beta\)-function, defined by
\begin{eqnarray}
 \beta_Q = \mu\frac{\D Q}{\D\mu}\;. 
\end{eqnarray}
Note that in equation (\ref{eq:BetaFunctionSusy}), for complex quantities 
$V_A$ we have to treat the complex conjugates  $V^*_A$ as 
independent variables. 
We use the notation \cite{Antusch:2001ck}

\begin{eqnarray}
\Braket{\frac{\D F}{\D x}|y} := 
 \left\{
 \begin{array}{ll}
        \displaystyle\frac{\D F}{\D x}y
        &\text{for scalars}\;x,y\\
        \displaystyle\sum_{n}\frac{\D F}{\D x_{n}} y_{n}
        \quad
        &\text{for vectors}\;x=(x_{m}),y=(y_{m})\\
        \displaystyle\sum_{m,n}\frac{\D F}{\D x_{mn}} y_{mn}
        \quad
        &\text{for matrices}\;x=(x_{mn}),y=(y_{mn})\\
        \dots
        &\text{etc.}
 \end{array}\right.
\end{eqnarray}
and $D_{V_A}$ is related to the mass dimension of $V_A$ as indicated in equation 
\eqref{eq:QBare}. 
Equation (\ref{eq:BetaFunctionSusy}) allows to compute the
$\beta$-functions directly from the wavefunction renormalization constants,
calculated with the supergraph technique\footnote{
Another way of calculating superpotential $\beta$-functions 
is based on
the superfield anomalous dimensions, which have been calculated 
to the 3-loop level in 
\cite{Jack:1996qq}.}. It has a form that  
can easily be used for computer algebra calculations.

\section{Applications}

\subsection{Two-Loop $\beta$-Functions in the MSSM Extended by Singlet
Superfields}\label{sec:MSSMExtendedBySinglets}

 We consider a supersymmetric model containing
 the same fields as the 
 MSSM and additionally the singlet ``neutrino'' superfield
 which we will denote by \(\SuperField{\nu}\). In order to obtain the
 $\beta$-functions for the Yukawa matrices and a mass matrix for the
 neutrino superfield, we will omit the soft 
 SUSY breaking terms\footnote{The calculation of the RGE's for the soft 
 SUSY breaking operators can be found in 
 \cite{Martin:1994zk,Yamada:1994id,Jack:1994kd}.},
 since they do
 not affect the considered $\beta$-functions above the scale of the soft
 supersymmetry breaking mass terms. 
 Threshold effects at low energy scales are e.g.\ 
 discussed in \cite{Chankowski:2001hx}.

Thus the Yukawa part of the superpotential is given by 
\begin{eqnarray}
 \mathscr{W}_\mathrm{Yukawa} & = &
 (Y_e)_{gf}\SuperField{e}^{\ChargeC g}
 	\SuperField{h}^{(1)}_a\varepsilon^{ab}\SuperField{l}^f_b
 +(Y_\nu)_{gf}\SuperField{\nu}^{\ChargeC g}
 	\SuperField{h}^{(2)}_a\varepsilon^{ab}\SuperField{l}^f_b
 \nonumber \\
 &&
 +(Y_d)_{gf}\SuperField{d}^{\ChargeC g}
 	\SuperField{h}^{(1)}_a\varepsilon^{ab}\SuperField{q}^f_b
 +(Y_u)_{gf}\SuperField{u}^{\ChargeC g}
 \SuperField{h}^{(2)}_a (\varepsilon^T)^{ab}  \SuperField{q}_b^f\; .	
\end{eqnarray}
The superfields \(\SuperField{e}^{\ChargeC }\), \(\SuperField{d}^{\ChargeC }\)
and \(\SuperField{u}^{\ChargeC}\) contain
the \(\mathrm{SU}(2)_\mathrm{L}\)-singlet charged leptons, 
down-type quarks 
and up-type quarks, respectively, 
and $\SuperField{q}$
contains the SU(2)$_\mathrm{L}$ quark doublets.
Their quantum numbers are specified in table \ref{tab:QuantumNumbersOfNuMSSM}.
In addition we consider a mass term for the singlet neutrino
superfield
\begin{eqnarray}\label{eq:BilinearCoupling}
 \mathscr{W}_\mathrm{Mass} 
 = \tfrac{1}{2}\,
 \SuperField{\nu}^{\mathrm{C}\,f} M^{}_{fg}
\SuperField{\nu}^{\mathrm{C}\,g}\; ,
\end{eqnarray}
that may be relevant for generating neutrino masses
in the see-saw scenario.

The quantum numbers of the superfields are listed in
table \ref{tab:QuantumNumbersOfNuMSSM}. Note that we use GUT 
charge normalization for the \(\mathrm{U}(1)_\mathrm{Y}\) charge.
\begin{table}[h]
\begin{center}
\begin{longtable}{l|cccccccc}
	Field &
		\(\SuperField{h}^{(1)}\) & \(\SuperField{h}^{(2)}\) 
		& \(\SuperField{q}\) 
			& \(\SFConjugate{\SuperField{d}}\) & \(\SFConjugate{\SuperField{u}}\) 
		& \(\SuperField{l}\) & \(\SFConjugate{\SuperField{e}}\) 
			& \(\SFConjugate{\SuperField{\nu}}\) \\
	\hline
	\(\vphantom{\bigg|}\sqrt{\frac{5}{3}}q_\mathrm{Y}\)
		& \(-\tfrac{1}{2}\) & \(+\tfrac{1}{2}\)
		& \(+\tfrac{1}{6}\)  & \(+\tfrac{1}{3}\) & \(-\tfrac{2}{3}\)
		& \(-\tfrac{1}{2}\) & \(+1\) & \(0\)  \\
	\(\vphantom{\big|}\mathrm{SU}(2)_\mathrm{L}\)	
		& \(\boldsymbol{2}\) & \(\boldsymbol{2}\)
		& \(\boldsymbol{2}\) & \(\boldsymbol{1}\) & \(\boldsymbol{1}\)
		& \(\boldsymbol{2}\) & \(\boldsymbol{1}\) & \(\boldsymbol{1}\) \\
	\(\vphantom{\big|}\mathrm{SU}(3)_\mathrm{C}\)	
		& \(\boldsymbol{1}\) & \(\boldsymbol{1}\) 
		& \(\boldsymbol{3}\) & \(\overline{\boldsymbol{3}}\) 
			& \(\overline{\boldsymbol{3}}\) 
		& \(\boldsymbol{1}\) & \(\boldsymbol{1}\) & \(\boldsymbol{1}\) 
\label{tab:QuantumNumbersOfNuMSSM}
\end{longtable}
\setcounter{table}{0}
\caption{Quantum numbers of the superfields. \(q_\mathrm{Y}\)
denotes the \(\mathrm{U}(1)_\mathrm{Y}\) charge in GUT normalization.}
\end{center}
\end{table}

Using equation \eqref{eq:OneLoopDeltaZ},
this leads to the \(1/\epsilon\)-coefficients of the 
wavefunction renormalization constants
\begin{subequations}\label{eq:ZMSSM1}
\begin{eqnarray}
 -(4\pi)^2\,Z_{\SuperField{h}^{(1)},1}^{(1)}
 & = &
 6\,\Tr(Y_d^\dagger\cdot Y_d) 
 + 2\,\Tr(Y_e^\dagger\cdot Y_e)
 - \frac{3}{5}\,g_1^2
 - 3\,g_2^2 
 \;,\\
 -(4\pi)^2\,Z_{\SuperField{h}^{(2)},1}^{(1)}
 & = &
 6\,\Tr(Y_u^\dagger\cdot Y_u)
 +  2\,\Tr(Y_\nu^\dagger\cdot Y_\nu) 
 - \frac{3}{5}\,g_1^2 
 - 3\,g_2^2 
 \;,\label{eq:Zh21}\\
 -(4\pi)^2\,Z_{\SuperField{q},1}^{(1)}
 & = &
 2\,Y_d^\dagger\cdot Y_d 
 + 2\,Y_u^\dagger\cdot Y_u 
 - \frac{1}{15}\,g_1^2
 - 3\,g_2^2 
 - \frac{16}{3}\,g_3^2
 \;,\\
 -(4\pi)^2\,Z_{\SuperField{d}^\ChargeC,1}^{(1)}
 & = &
 4\,Y_d^*\cdot Y_d^T 
 - \frac{4}{15}\,g_1^2
 - \frac{16}{3}\,g_3^2
 \;,\\
 -(4\pi)^2\,Z_{\SuperField{u}^\ChargeC,1}^{(1)}
 & = &
 4\,Y_u^*\cdot Y_u^T 
 - \frac{16}{15}\,g_1^2
 - \frac{16}{3}\,g_3^2
 \;,\\
 -(4\pi)^2\,Z_{\SuperField{l},1}^{(1)}
 & = &
 2\,Y_e^\dagger\cdot Y_e 
 + 2\,Y_\nu^\dagger\cdot Y_\nu 
 - \frac{3}{5}\,g_1^2
 -  3\,g_2^2
 \;,\label{eq:Zl1}\\
 -(4\pi)^2\,Z_{\SuperField{e}^\ChargeC,1}^{(1)}
 & = &
 4\,Y_e^*\cdot Y_e^T 
 - \frac{12}{5}\,g_1^2
 \;,\\
 -(4\pi)^2\,Z_{\SuperField{\nu}^\ChargeC,1}^{(1)}
 & = &
 4\,Y_\nu^*\cdot Y_\nu^T
 \;,
\end{eqnarray}
\end{subequations}
where the \(Y\)'s as well as the last six \(Z\)-factors 
are of course matrices in flavour space.
From the two-loop diagrams we get
\begin{subequations}\label{eq:ZMSSM2}
\begin{eqnarray}
 -(4\pi)^4\,\frac{Z_{\SuperField{h}^{(1)},1}^{(2)}}{2}
 & = &
 - 9\,\Tr(Y_d\cdot Y_d^\dagger\cdot Y_d\cdot Y_d^\dagger) 
 - 3\,\Tr(Y_u\cdot Y_d^\dagger\cdot Y_d\cdot Y_u^\dagger) 
 \nonumber\\
 & &
 - 3\,\Tr(Y_e\cdot Y_e^\dagger\cdot Y_e\cdot Y_e^\dagger) 
 - \Tr(Y_e\cdot Y_\nu^\dagger\cdot Y_\nu\cdot Y_e^\dagger)
 \nonumber\\
 & &
 - \frac{2}{5}\,g_1^2\,\Tr(Y_d^\dagger\cdot Y_d)
 + \frac{6}{5}\,g_1^2\,\Tr(Y_e^\dagger\cdot Y_e) 
 + 16\,g_3^2\,\Tr(Y_d^\dagger\cdot Y_d) 
 \nonumber\\
 & &
 + \frac{207}{100}\,g_1^4
 + \frac{9}{10}\,g_1^2\,g_2^2
 + \frac{15}{4}\,g_2^4\;,
\end{eqnarray}
\begin{eqnarray}
 -(4\pi)^4\,\frac{Z_{\SuperField{h}^{(2)},1}^{(2)}}{2}\label{eq:TwoLoopZh2}
 & = &
 - 3\,\Tr(Y_u\cdot Y_d^\dagger\cdot Y_d\cdot Y_u^\dagger) 
 - 9\,\Tr(Y_u\cdot Y_u^\dagger\cdot Y_u\cdot Y_u^\dagger)
 \nonumber\\
 & &
 - \Tr(Y_\nu\cdot Y_e^\dagger\cdot Y_e\cdot Y_\nu^\dagger) 
 - 3\,\Tr(Y_\nu\cdot Y_\nu^\dagger\cdot Y_\nu\cdot Y_\nu^\dagger) 
 \nonumber\\
 & &
 + \frac{4}{5}\,g_1^2\,\Tr(Y_u^\dagger\cdot Y_u) 
 + 16\,g_3^2\,\Tr(Y_u^\dagger\cdot Y_u) 
 \nonumber\\
 & &
 + \frac{207}{100}\,g_1^4 
 + \frac{9}{10}\,g_1^2\,g_2^2 
 + \frac{15}{4}\,g_2^4\;,
\end{eqnarray}
\begin{eqnarray}
 -(4\pi)^4\,\frac{Z_{\SuperField{q},1}^{(2)}}{2}
 & = &
 - 2\,Y_d^\dagger\cdot Y_d\cdot Y_d^\dagger\cdot Y_d 
 - 2\,Y_u^\dagger\cdot Y_u\cdot Y_u^\dagger\cdot Y_u 
 \nonumber\\
 & &
 - 3\,Y_d^\dagger\cdot Y_d\,\Tr(Y_d\cdot Y_d^\dagger) 
 -  3\,Y_u^\dagger\cdot Y_u\,\Tr(Y_u\cdot Y_u^\dagger)
 \nonumber\\
 & &
 -  Y_d^\dagger\cdot Y_d\,\Tr(Y_e\cdot Y_e^\dagger) 
 -  Y_u^\dagger\cdot Y_u\,\Tr(Y_\nu\cdot Y_\nu^\dagger) 
 \nonumber\\
 & & 
 + \frac{2}{5}\,g_1^2\,Y_d^\dagger\cdot Y_d
 + \frac{4}{5}\,g_1^2\,Y_u^\dagger\cdot Y_u
 \nonumber\\
 & &
 + \frac{199}{900}\,g_1^4
 + \frac{1}{10}\,g_1^2\,g_2^2
 + \frac{15}{4}\,g_2^4
 \nonumber\\
 & &
 + \frac{8}{45}\,g_1^2\,g_3^2 
 + 8\,g_2^2\,g_3^2 
 - \frac{8}{9}\,g_3^4\;,
\end{eqnarray}
\begin{eqnarray}
 -(4\pi)^4\,\frac{Z_{\SuperField{d}^\ChargeC,1}^{(2)}}{2}
 & = &
 - 2\,Y_d^*\cdot Y_d^T\cdot Y_d^*\cdot Y_d^T 
 - 2\,Y_d^*\cdot Y_u^T\cdot Y_u^*\cdot Y_d^T 
 \nonumber\\
 & &
 - 6\,Y_d^*\cdot Y_d^T\,\Tr(Y_d\cdot Y_d^\dagger) 
 - 2\,Y_d^*\cdot Y_d^T\,\Tr(Y_e\cdot Y_e^\dagger)
 \nonumber\\
 & &
 + \frac{2}{5}\,g_1^2\,Y_d^*\cdot Y_d^T
 + 6\,g_2^2\,Y_d^*\cdot Y_d^T
 \nonumber\\
 & &
 + \frac{202}{225}\,g_1^4 
 + \frac{32}{45}\,g_1^2\,g_3^2 
 - \frac{8}{9}\,g_3^4 \;,
\end{eqnarray}
\begin{eqnarray}
 -(4\pi)^4\,\frac{Z_{\SuperField{u}^\ChargeC,1}^{(2)}}{2}
 & = &
 - 2\,Y_u^*\cdot Y_d^T\cdot Y_d^*\cdot Y_u^T 
 - 2\,Y_u^*\cdot Y_u^T\cdot Y_u^*\cdot Y_u^T 
 \nonumber\\
 & &
 - 2\,Y_u^*\cdot Y_u^T\,\Tr(Y_\nu\cdot Y_\nu^\dagger) 
 - 6\,Y_u^*\cdot Y_u^T\,\Tr(Y_u\cdot Y_u^\dagger)
 \nonumber\\
 & &
 - \frac{2}{5}\,g_1^2\,Y_u^*\cdot Y_u^T
 + 6\,g_2^2 \,Y_u^*\cdot Y_u^T
 \nonumber\\
 & &
 + \frac{856}{225}\,g_1^4 
 + \frac{128}{45}\,g_1^2\,g_3^2 
 - \frac{8}{9}\,g_3^4\;,
\end{eqnarray}
\begin{eqnarray}
 -(4\pi)^4\,\frac{Z_{\SuperField{l},1}^{(2)}}{2}\label{eq:TwoLoopZl}
 & = &
 - 2\,Y_e^\dagger\cdot Y_e\cdot Y_e^\dagger\cdot Y_e 
 - 2\,Y_\nu^\dagger\cdot Y_\nu\cdot Y_\nu^\dagger\cdot Y_\nu 
 \nonumber\\
 & &
 - 3\,Y_e^\dagger\cdot Y_e\,\Tr(Y_d\cdot Y_d^\dagger) 
 - Y_e^\dagger\cdot Y_e\,\Tr(Y_e\cdot Y_e^\dagger) 
 \nonumber\\
 & &
 - 3\,Y_\nu^\dagger\cdot Y_\nu\,\Tr(Y_u\cdot Y_u^\dagger)
 - Y_\nu^\dagger\cdot Y_\nu\,\Tr(Y_\nu\cdot Y_\nu^\dagger) 
 \nonumber\\
 & &
 + \frac{6}{5}\,g_1^2\,Y_e^\dagger\cdot Y_e 
 + \frac{207}{100} \,g_1^4
 + \frac{9}{10}\,g_1^2\,g_2^2 
 + \frac{15}{4}\,g_2^4\;,
\end{eqnarray}
\begin{eqnarray}
 -(4\pi)^4\,\frac{Z_{\SuperField{e}^\ChargeC,1}^{(2)}}{2}
 & = &
 - 2\,Y_e^*\cdot Y_e^T\cdot Y_e^*\cdot Y_e^T 
 - 2 \,Y_e^*\cdot Y_\nu^T\cdot Y_\nu^*\cdot Y_e^T 
 \nonumber\\
 & &
 - 6\,Y_e^*\cdot Y_e^T\,\Tr(Y_d\cdot Y_d^\dagger) 
 - 2\,Y_e^*\cdot Y_e^T\,\Tr(Y_e\cdot Y_e^\dagger)
 \nonumber\\
 & &
 - \frac{6}{5}\,g_1^2\,Y_e^*\cdot Y_e^T 
 + 6\,g_2^2\,Y_e^*\cdot Y_e^T
 + \frac{234}{25}\,g_1^4\;,
\end{eqnarray}
\begin{eqnarray}
 -(4\pi)^4\,\frac{Z_{\SuperField{\nu}^\ChargeC,1}^{(2)}}{2}
 & = &
 - 2\,Y_\nu^*\cdot Y_e^T\cdot Y_e^*\cdot Y_\nu^T 
 - 2\,Y_\nu^*\cdot Y_\nu^T\cdot Y_\nu^*\cdot Y_\nu^T 
 \nonumber\\
 & &
 - 6\,Y_\nu^*\cdot Y_\nu^T\,\Tr(Y_u\cdot Y_u^\dagger)
 - 2\,Y_\nu^*\cdot Y_\nu^T\,\Tr(Y_\nu\cdot Y_\nu^\dagger) 
 \nonumber\\
 & &
 + \frac{6}{5}\,g_1^2\,Y_\nu^*\cdot Y_\nu^T
 + 6\,g_2^2\,Y_\nu^*\cdot Y_\nu^T\;,
\end{eqnarray}
\end{subequations}
respectively.
From these, the two-loop Yukawa RGE's are derived,
\begin{equation}
 \mu\frac{\D Y_x}{\D \mu}
 =
 \frac{1}{(4\pi)^2}\beta_{Y_x}^{(1)}+\frac{1}{(4\pi)^4}\beta_{Y_x}^{(2)}\;, 
\end{equation}
where \(x\in\{d,u,e,\nu\}\).
Using equation (\ref{eq:BetaFunctionSusy}), the one-loop contributions 
to the \(\beta\)-functions are given by
\begin{subequations}
\begin{eqnarray}
 \beta_{Y_d}^{(1)} 
 & = & Y_d\cdot\left\{\vphantom{\frac{1}{2}}\right.
 3\,Y_d^\dagger\cdot Y_d 
 + Y_u^\dagger\cdot Y_u 
 + 3\,\Tr(Y_d^\dagger\cdot Y_d) 
 + \Tr(Y_e^\dagger\cdot Y_e)
 \nonumber\\
 & & \hphantom{Y_d\cdot\left\{\right.}
 {}
 - \frac{7}{15}\,g_1^2
 - 3\,g_2^2 
 - \frac{16}{3}\,g_3^2
 \left.\vphantom{\frac{1}{2}}\right\}\;,
\end{eqnarray}
\begin{eqnarray}
 \beta_{Y_u}^{(1)} 
 & = & Y_u\cdot\left\{\vphantom{\frac{1}{2}}\right.
 Y_d^\dagger\cdot Y_d 
 + 3\, Y_u^\dagger\cdot Y_u 
 + \Tr(Y_\nu^\dagger\cdot Y_\nu) 
 + 3\,\Tr(Y_u^\dagger\cdot Y_u)
 \nonumber\\
 & & \hphantom{Y_u\cdot\left\{\right.}
 {}
 - \frac{13}{15}\,g_1^2 
 - 3\,g_2^2 
 - \frac{16}{3}\,g_3^2 
 \left.\vphantom{\frac{1}{2}}\right\}\;
\end{eqnarray}
\begin{eqnarray}
\beta_{Y_e}^{(1)} 
 & = & Y_e\cdot\left\{\vphantom{\frac{1}{2}}\right.
 3\,Y_e^\dagger\cdot Y_e 
 + Y_\nu^\dagger\cdot Y_\nu 
 + 3\,\Tr(Y_d^\dagger\cdot Y_d) 
 + \Tr(Y_e^\dagger\cdot Y_e)
 \nonumber\\
 & & \hphantom{Y_e\cdot\left\{\right.}
 {}- \frac{9}{5}\,g_1^2 
 - 3\,g_2^2 
 \left.\vphantom{\frac{1}{2}}\right\}\;,
\end{eqnarray}
\begin{eqnarray} \label{eq:MSSMBeta2LoopYnu}
\beta_{Y_\nu}^{(1)} 
& = & Y_\nu\cdot\left\{\vphantom{\frac{1}{2}}\right.
 Y_e^\dagger\cdot Y_e 
 + 3\,Y_\nu^\dagger\cdot Y_\nu 
 + 3\,\Tr(Y_u^\dagger\cdot Y_u)
 + \Tr(Y_\nu^\dagger\cdot Y_\nu) 
 \nonumber\\
 & & \hphantom{Y_\nu\cdot\left\{\right.}
 {}- \frac{3}{5}\,g_1^2 
 - 3\,g_2^2 
 \left.\vphantom{\frac{1}{2}}\right\}\;,
\end{eqnarray}
\end{subequations}
and the two-loop contributions are
\begin{subequations}
\begin{eqnarray}
 \beta_{Y_d}^{(2)}
 & = &
 Y_d\cdot\left\{\vphantom{\frac{1}{2}}
 - 4\,Y_d^\dagger\cdot Y_d\cdot Y_d^\dagger\cdot Y_d 
 - 2\,Y_u^\dagger\cdot Y_u\cdot Y_d^\dagger\cdot Y_d 
 - 2\,Y_u^\dagger\cdot Y_u\cdot Y_u^\dagger\cdot Y_u
 \right.
 \nonumber\\
 & & \hphantom{Y_d\cdot\left\{\right.}
 {}- 9\,\Tr(Y_d\cdot Y_d^\dagger\cdot Y_d\cdot Y_d^\dagger)
 - 3\,\Tr(Y_d\cdot Y_u^\dagger\cdot Y_u\cdot Y_d^\dagger) 
 \nonumber\\
 & & \hphantom{Y_d\cdot\left\{\right.}
 {}- 3\,\Tr(Y_e\cdot Y_e^\dagger\cdot Y_e\cdot Y_e^\dagger) 
 - \Tr(Y_e\cdot Y_\nu^\dagger\cdot Y_\nu\cdot Y_e^\dagger) 
 \nonumber\\
 & & \hphantom{Y_d\cdot\left\{\right.}
 {}- 9\,Y_d^\dagger\cdot Y_d\,\Tr(Y_d\cdot Y_d^\dagger) 
 - 3\,Y_d^\dagger\cdot Y_d\,\Tr(Y_e\cdot Y_e^\dagger) 
 \nonumber\\
 & & \hphantom{Y_d\cdot\left\{\right.}
 {}- Y_u^\dagger\cdot Y_u\,\Tr(Y_\nu\cdot 
 	Y_\nu^\dagger) 
 - 3\, Y_u^\dagger\cdot Y_u\,\Tr(Y_u\cdot Y_u^\dagger)
 \nonumber\\
 & & \hphantom{Y_d\cdot\left\{\right.}
 {}+ 6\,g_2^2\,Y_d^\dagger\cdot Y_d 
 + \frac{4}{5}\,g_1^2\, Y_d^\dagger\cdot Y_d
 + \frac{4}{5}\,g_1^2\, Y_u^\dagger\cdot Y_u
 \nonumber\\
 & & \hphantom{Y_d\cdot\left\{\right.}
 {}- \frac{2}{5}\,g_1^2\,\Tr(Y_d^\dagger\cdot Y_d)
 + \frac{6}{5}\,g_1^2\,\Tr(Y_e^\dagger\cdot Y_e) 
 + 16\,g_3^2\, \Tr(Y_d^\dagger\cdot Y_d) 
 \nonumber\\
 & & \hphantom{Y_d\cdot\left\{\right.}
 {}+ \frac{287}{90}\,g_1^4 
 + g_1^2\,g_2^2 
 + \frac{15}{2}\,g_2^4 
 + \frac{8}{9}\,g_1^2\,g_3^2 
 + 8\,g_2^2\,g_3^2 
 - \frac{16}{9}\,g_3^4
 \left.\vphantom{\frac{1}{2}}\right\}\;,
\end{eqnarray}\begin{eqnarray}
 \beta_{Y_u}^{(2)}
 & = &
 Y_u\cdot\left\{
 - 2\,Y_d^\dagger\cdot Y_d\cdot Y_d^\dagger\cdot Y_d
 - 2\,Y_d^\dagger\cdot Y_d\cdot Y_u^\dagger\cdot Y_u 
 - 4\,Y_u^\dagger\cdot Y_u\cdot Y_u^\dagger\cdot Y_u 
 \right.
 \nonumber\\
 & & \hphantom{Y_u\cdot\left\{\right.}
 {}- 3\,Y_d^\dagger\cdot Y_d\,\Tr(Y_d\cdot Y_d^\dagger)
 - Y_d^\dagger\cdot Y_d\,\Tr(Y_e\cdot Y_e^\dagger) 
 \nonumber\\
 & & \hphantom{Y_u\cdot\left\{\right.}
 {}- 9\,Y_u^\dagger\cdot Y_u\,\Tr(Y_u\cdot Y_u^\dagger) 
 - 3\,Y_u^\dagger\cdot Y_u\,\Tr(Y_\nu\cdot Y_\nu^\dagger) 
 \nonumber\\
 & & \hphantom{Y_u\cdot\left\{\right.}
 {}- 3\,\Tr(Y_u\cdot Y_d^\dagger\cdot Y_d\cdot Y_u^\dagger) 
 - 9\,\Tr(Y_u\cdot Y_u^\dagger\cdot Y_u\cdot Y_u^\dagger) 
 \nonumber\\
 & & \hphantom{Y_u\cdot\left\{\right.}
 {}- \Tr(Y_\nu\cdot Y_e^\dagger\cdot Y_e\cdot Y_\nu^\dagger) 
 - 3\,\Tr(Y_\nu\cdot Y_\nu^\dagger\cdot Y_\nu\cdot Y_\nu^\dagger) 
 \nonumber\\
 & & \hphantom{Y_u\cdot\left\{\right.}
 {}+ \frac{2}{5}\,g_1^2\, Y_d^\dagger\cdot Y_d 
 + \frac{2}{5}\,g_1^2\, Y_u^\dagger\cdot Y_u
 + 6\,g_2^2 \,Y_u^\dagger\cdot Y_u
 \nonumber\\
 & & \hphantom{Y_u\cdot\left\{\right.}
 {}+ \frac{4}{5}\,g_1^2\,\Tr(Y_u^\dagger\cdot Y_u) 
 + 16\,g_3^2\,\Tr(Y_u^\dagger\cdot Y_u)
 + \frac{2743}{450}\,g_1^4
 \nonumber\\
 & & \hphantom{Y_u\cdot\left\{\right.}
 {}+ g_1^2\,g_2^2 
 + \frac{15}{2}\,g_2^4
 + \frac{136}{45}\,g_1^2\,g_3^2
 + 8\,g_2^2\,g_3^2 
 - \frac{16}{9}\,g_3^4
 \left.\vphantom{\frac{1}{2}}\right\}\;,
\end{eqnarray}\begin{eqnarray}
  \beta_{Y_e}^{(2)}
 & = &
 Y_e\cdot\left\{\vphantom{\frac{1}{2}}\right.
 - 4\,Y_e^\dagger\cdot Y_e\cdot Y_e^\dagger\cdot Y_e 
 - 2\,Y_\nu^\dagger\cdot Y_\nu
 	\cdot Y_e^\dagger\cdot Y_e 
 - 2\,Y_\nu^\dagger\cdot Y_\nu
 	\cdot Y_\nu^\dagger\cdot Y_\nu 
 \nonumber\\
 & & \hphantom{Y_e\cdot\left\{\right.}
 {}- 9\,Y_e^\dagger\cdot Y_e\,\Tr(Y_d\cdot Y_d^\dagger) 
 - 3\,Y_e^\dagger\cdot Y_e\,\Tr(Y_e\cdot Y_e^\dagger) 
 \nonumber\\
 & & \hphantom{Y_e\cdot\left\{\right.}
 {}- Y_\nu^\dagger\cdot Y_\nu\,\Tr(Y_\nu\cdot Y_\nu^\dagger) 
 - 3\,Y_\nu^\dagger\cdot Y_\nu\,\Tr(Y_u\cdot Y_u^\dagger) 
 \nonumber\\
 & & \hphantom{Y_e\cdot\left\{\right.}
 {}- 9\,\Tr(Y_d\cdot Y_d^\dagger\cdot Y_d\cdot Y_d^\dagger) 
 - 3\,\Tr(Y_d\cdot Y_u^\dagger\cdot Y_u\cdot Y_d^\dagger) 
 \nonumber\\
 & & \hphantom{Y_e\cdot\left\{\right.}
 {}- 3\,\Tr(Y_e\cdot Y_e^\dagger\cdot Y_e\cdot Y_e^\dagger) 
 - \Tr(Y_e\cdot Y_\nu^\dagger\cdot Y_\nu\cdot Y_e^\dagger) 
 + \frac{6}{5}\,g_1^2\,\Tr(Y_e^\dagger\cdot Y_e)
 \nonumber\\
 & & \hphantom{Y_e\cdot\left\{\right.}
 {}+ 6\,g_2^2\,Y_e^\dagger\cdot Y_e
 - \frac{2}{5}\,g_1^2\,\Tr(Y_d^\dagger\cdot Y_d) 
 + 16\,g_3^2\, \Tr(Y_d^\dagger\cdot Y_d) 
 \nonumber\\
 & & \hphantom{Y_e\cdot\left\{\right.}
 {}+ \frac{27}{2}\,g_1^4
 + \frac{9}{5}\,g_1^2\,g_2^2
 + \frac{15}{2}\,g_2^4
 \left.\vphantom{\frac{1}{2}}\right\}\;,
\end{eqnarray}\begin{eqnarray}
 \beta_{Y_\nu}^{(2)}
 & = &
 Y_\nu\cdot\left\{\vphantom{\frac{1}{2}}\right.
 - 2\,Y_e^\dagger\cdot Y_e\cdot Y_e^\dagger\cdot Y_e 
 - 2\,Y_e^\dagger\cdot Y_e\cdot Y_\nu^\dagger
 	\cdot Y_\nu 
 - 4\,Y_\nu^\dagger\cdot Y_\nu
 	\cdot Y_\nu^\dagger\cdot Y_\nu 
 \nonumber\\
 & & \hphantom{Y_\nu\cdot\left\{\right.}
 {}- 3\,Y_e^\dagger\cdot Y_e\,\Tr(Y_d\cdot Y_d^\dagger) 
 - Y_e^\dagger\cdot Y_e\,\Tr(Y_e\cdot Y_e^\dagger) 
 \nonumber\\
 & & \hphantom{Y_\nu\cdot\left\{\right.}
 - 3\,Y_\nu^\dagger\cdot Y_\nu\,
   \Tr(Y_\nu\cdot Y_\nu^\dagger) 
 - 9\,Y_\nu^\dagger\cdot Y_\nu\,
  	\Tr(Y_u\cdot Y_u^\dagger) 
 \nonumber\\
 & & \hphantom{Y_\nu\cdot\left\{\right.}
 {}- \Tr(Y_\nu\cdot Y_e^\dagger\cdot Y_e\cdot Y_\nu^\dagger)
 - 3\,\Tr(Y_\nu^\dagger\cdot Y_\nu\cdot 
 	Y_\nu^\dagger\cdot  Y_\nu) 
 \nonumber\\
 & & \hphantom{Y_\nu\cdot\left\{\right.}
 {}- 3\,\Tr(Y_u\cdot Y_d^\dagger\cdot Y_d\cdot Y_u^\dagger) 
 - 9\,\Tr(Y_u\cdot Y_u^\dagger\cdot Y_u\cdot Y_u^\dagger) 
 \nonumber\\
 & & \hphantom{Y_\nu\cdot\left\{\right.}
 {}+ \frac{6}{5}\,g_1^2\,Y_e^\dagger\cdot Y_e
 + \frac{6}{5}\,g_1^2\,Y_\nu^\dagger\cdot  Y_\nu
 + 6\,g_2^2\,Y_\nu^\dagger\cdot Y_\nu
 \nonumber\\
 & & \hphantom{Y_\nu\cdot\left\{\right.}
 {}+ \frac{4}{5}\,g_1^2\, \Tr(Y_u^\dagger\cdot Y_u)
 + 16\,g_3^2\,\Tr(Y_u^\dagger\cdot Y_u)
 \nonumber\\
 & & \hphantom{Y_\nu\cdot\left\{\right.}
 {}+ \frac{207}{50}\,g_1^4
 + \frac{9}{5}\,g_1^2\,g_2^2
 + \frac{15}{2}\,g_2^4
 \left.\vphantom{\frac{1}{2}}\right\}\;.
\end{eqnarray}
\end{subequations}
Note that the two-loop MSSM RGE's for \(Y_d\), \(Y_u\) and \(Y_e\) are 
easily obtained by setting \(Y_\nu=0\). The effort is clearly reduced compared
to component field calculations 
\cite{Machacek:1983tz,Machacek:1983fi,Machacek:1984zw}.

\subsection{Two-Loop $\beta$-Function for the Effective Neutrino Mass Operator}

We now apply our method to calculate the $\beta$-function for the lowest
dimensional effective 
neutrino mass operator, which is contained in the $F$-term of  the superpotential
\begin{eqnarray}\label{eq:EffNuMassOperator}
 \mathscr{W}_{\kappa}^{\mathrm{MSSM}} 
 =-\frac{1}{4} 
  {\kappa}^{}_{gf} \,\SuperField{l}^{g}_c\varepsilon^{cd}
 \SuperField{h}^{(2)}_d\, 
 \, \SuperField{l}_{b}^{f}\varepsilon^{ba} \SuperField{h}^{(2)}_a 
 +\text{h.c.} \; .
\end{eqnarray}
It can e.g.\ be obtained by integrating out the singlet superfield 
\(\SuperField{\nu}\) of the
model described in section \ref{sec:MSSMExtendedBySinglets} 
at leading order in the effective theory.
The $\beta$-function can easily be computed using our method. 
Substituting \(D_{g_i}=D_{Y_x}=\frac{1}{2}\) with
\(i\in \{1,2,3\}\) and \(x \in \{u,d,e\}\), we get from equation 
(\ref{eq:BetaFunctionSusy}) 
\begin{eqnarray}
 \beta_\kappa = -Z_{\SuperField{h}^{(2)},1}\cdot \kappa 
 - \frac{1}{2}\,Z_{\SuperField{l},1}^T\cdot \kappa
 - \frac{1}{2}\,\kappa\cdot Z_{\SuperField{l},1}\;.
\end{eqnarray}
We can thus write the $\beta$-function for \(\kappa\) in the form
\begin{eqnarray}
 \beta_\kappa= X^T\cdot\kappa+\kappa\cdot X +\alpha\,\kappa \; ,
\end{eqnarray}
where the complete flavour diagonal part is contained in $\alpha$.
We further split $X=X^{(1)}+X^{(2)}$ and $\alpha=\alpha^{(1)}+\alpha^{(2)}$
into their one loop and two loop part. 
Plugging in the wavefunction renormalization constants of equation 
\eqref{eq:Zh21} and \eqref{eq:Zl1} and setting \(Y_\nu=0\), 
our method reproduces the one loop results
of \cite{Chankowski:1993tx,Babu:1993qv,Antusch:2002vn}
\begin{subequations}
\begin{eqnarray}
 (4\pi)^2\,X^{(1)} 
 & = & 
 Y_e^\dagger\cdot Y_e\;,\\
 (4\pi)^2\,\alpha^{(1)} 
 & = &
 - \frac{6}{5}\,g_1^2 
 - 6\,g_2^2  
 + 6\,\Tr(Y_u^\dagger\cdot Y_u)\;.
\end{eqnarray}
\end{subequations}
Note that for \(\mathrm{U}(1)_\mathrm{Y}\), we use GUT charge 
normalization as specified in table \ref{tab:QuantumNumbersOfNuMSSM}.
At two-loop, with the 
wavefunction renormalization constants given in  
equations (\ref{eq:TwoLoopZh2}) and (\ref{eq:TwoLoopZl}), we obtain
\begin{eqnarray}
 (4\pi)^4\,X^{(2)}  
 & = &
 - 2\, Y_e^\dagger\cdot Y_e\cdot Y_e^\dagger\cdot Y_e 
 \nonumber\\
 & &
 {}+\left(
  \frac{6}{5}\,g_1^2
  - \Tr(Y_e\cdot Y_e^\dagger) 
  - 3\,\Tr(Y_d\cdot Y_d^\dagger)
 \right)
 \,Y_e^\dagger\cdot Y_e
\end{eqnarray}
and
\begin{eqnarray}
 (4\pi)^4\,\alpha^{(2)}
 & = &
 -6\,\Tr(Y_u\cdot Y_d^\dagger\cdot Y_d\cdot Y_u^\dagger) 
 - 18\,\Tr(Y_u\cdot Y_u^\dagger\cdot Y_u\cdot Y_u^\dagger) 
 \nonumber
 \\
 & & 
 {}+ \frac{8}{5}\,g_1^2\,\Tr(Y_u^\dagger\cdot Y_u)
 + 32\,g_3^2\,\Tr(Y_u^\dagger\cdot Y_u)
 \nonumber\\
 & &
 {}+ \frac{207}{25}\,\,g_1^4
 + \frac{18}{5}\,g_1^2\,g_2^2 
 + 15\,g_2^4 \;.
\end{eqnarray}

\subsection{Two-Loop $\beta$-Function for the Mass of the Singlet Superfield}

From the wavefunction renormalization constants of the MSSM extended by singlet
superfields given in section 
\ref{sec:MSSMExtendedBySinglets}, the $\beta$-function for the 
bilinear coupling  of equation \eqref{eq:BilinearCoupling} 
can easily be computed using the 
formula of equation (\ref{eq:BetaFunctionSusy}).
At one-loop, we find
\begin{eqnarray}
 (4\pi)^2\beta_M^{(1)}=
 2\,M\cdot Y_\nu^*\cdot    Y_\nu^T 
  +  2\,Y_\nu\cdot Y_\nu^\dagger\cdot M
\end{eqnarray}
and the two-loop part of the $\beta$-function is given by
\begin{eqnarray}
 (4\pi)^4\beta_M^{(2)} 
 & = &
 M\cdot \left[\vphantom{\frac{1}{2}} \right.
 - 2\,Y_\nu^*\cdot    Y_e^T\cdot    Y_e^*\cdot    Y_\nu^T 
 - 2\,Y_\nu^*\cdot    Y_\nu^T\cdot    Y_\nu^*\cdot    Y_\nu^T 
 \nonumber\\
 && \hphantom{M\cdot \left[\right.}
 {}- 6\,Y_\nu^*\cdot    Y_\nu^T\, \Tr(Y_u\cdot     Y_u^\dagger) 
 - 2\,Y_\nu^*\cdot    Y_\nu^T\,   \Tr(Y_\nu\cdot     Y_\nu^\dagger) 
 \nonumber\\
 && \hphantom{M\cdot \left[\right.}
 {}+ \frac{6}{5}\,g_1^2\, Y_\nu^*\cdot Y_\nu^T
 + 6\,g_2^2\,Y_\nu^*\cdot    Y_\nu^T 
 \left.\vphantom{\frac{1}{2}} \right]
 \nonumber\\
 &&
 {}+\left[\vphantom{\frac{1}{2}} \right.
 - 2\,Y_\nu\cdot Y_e^\dagger\cdot Y_e\cdot    Y_\nu^\dagger
 - 2\,Y_\nu\cdot    Y_\nu^\dagger\cdot    Y_\nu\cdot Y_\nu^\dagger
 \nonumber\\
 &&\hphantom{+\left[\right.}
 {}- 6\,Y_\nu\cdot Y_\nu^\dagger\, \Tr(Y_u\cdot     Y_u^\dagger)
 - 2\,Y_\nu\cdot Y_\nu^\dagger\, \Tr(Y_\nu\cdot     Y_\nu^\dagger) 
 \nonumber\\
 &&\hphantom{+\left[\right.}
 {}+ \frac{6}{5}\,g_1^2\,Y_\nu\cdot Y_\nu^\dagger
 + 6\,g_2^2\,Y_\nu\cdot Y_\nu^\dagger 
 \left.\vphantom{\frac{1}{2}}\right]\cdot M\;.
\end{eqnarray}
In typical models of neutrino masses based on the see-saw mechanism,
the effective 
neutrino mass operator of equation (\ref{eq:EffNuMassOperator})
is obtained by integrating out the singlet superfields, which leads to the
relation 
$\kappa = 2  Y_\nu^T M^{-1} Y_\nu$. 
The $\beta$-function for $M$, together with the 
\(\beta\)-function for $Y_\nu$ of equation (\ref{eq:MSSMBeta2LoopYnu}),
 is therefore 
required to evolve the neutrino mass matrix
from the GUT scale to the scale of 
the largest eigenvalue of $M$, if the singlets have a direct mass term.

\section{Discussion and Conclusions}
We have presented a general method to calculate 
two loop \(\beta\)-functions
for renormalizable and non-renormalizable operators 
of the superpotential using superfield techniques. 
This method is very useful for model building since it provides a
construction kit which allows to calculate
the $\beta$-functions in a given supersymmetric GUT model 
with little effort.
We have applied this method to calculate the two-loop beta functions for the
lowest-dimensional effective neutrino mass operator in the 
MSSM and for the the Yukawa couplings and the mass matrix 
in the MSSM extended by singlet chiral superfields. 
We have computed and specified the wavefunction 
renormalization constants in the latter model, from which, using our method, 
the two loop RGE's for every, even higher dimensional, operator of the
superpotential can 
directly be computed. 
A classification of several higher-dimensional operators 
for generating neutrino Majorana masses is e.g.\ 
given in \cite{Babu:2001ex}. 
Many of them can be generalized for supersymmetric models. 
Their \(\beta\)-functions can easily be obtained with our method. 
The two loop \(\beta\)-function for the
lowest-dimensional effective neutrino mass operator is required
to increase the accuracy of many studies based on the  
one loop RGE \cite{Chankowski:1993tx,Babu:1993qv,Antusch:2002vn}, 
e.g.\
\cite{Ellis:1999my,Ibarra:1999se,Casas:1999tg,Balaji:2000gd,Balaji:2000ma,Kuo:2001ha,Froggatt:2002tb,Babu:2002ex,Dutta:2002pi}. 
This accuracy may be needed for the neutrino sector since 
due to the absence of hadronic uncertainties, high precision measurements of the
neutrino parameters may be achieved in future experiments.

\section*{Acknowledgments}

We would like to thank M.~Drees and M.~Lindner 
for useful discussions and M.~Schmidt for pointing out 
an error in some two-loop beta-functions. This work was supported by the 
``Sonderforschungsbereich~375 f\"ur Astro-Teilchenphysik der 
Deutschen Forschungsgemeinschaft''.
M.\ R.\ was partly supported by the 
``Promotionsstipendium des Freistaats Bayern''.

\end{document}